\documentclass[a4paper,11pt]{article}
\pdfoutput=1 

\usepackage{jheppub} 

\usepackage{tikz}
\newcommand*{\circled}[1]{\lower.7ex\hbox{\tikz\draw (0pt, 0pt)%
    circle (.5em) node {\makebox[1em][c]{\small #1}};}}

\title{Effect of quantum deformed black hole on BH shadow in two-dimensional Dilaton gravity}

\author[a,c]{Zhaoyi Xu}
\author[a]{Meirong Tang}

\affiliation[a]{College of Physics, Guizhou University, Guiyang 550025, China}
\affiliation[c]{Key Laboratory of Particle Astrophysics,
Institute of High Energy Physics, Chinese Academy of Sciences,
Beijing 100049, China}

\emailAdd{mrtang@ynao.ac.cn}

\abstract{In recent years, the study of quantum effects near the event horizon of black hole (BH) has attracted extensive attention. It has become one of the important methods to explore BH quantum properties by using the related properties of the quantum deformed black hole. In this work, we study the effect of quantum deformed black hole on BH shadow in two-dimensional Dilaton gravity. In this model, quantum effects are reflected on the quantum correction parameter m. By calculation, we find that: (1) the shape of the shadow boundary of a rotating black hole is determined by the BH spin $a$, the quantum correction parameter $m$ and the BH type parameter $n$; (2) when the spin $a=0$, the shape of the BH shadow is a perfect circle; when $a\neq 0$, the shape is distorted; if the quantum correction parameter $m=0$, their shapes reduce to the cases of Schwarzschild BH and Kerr BH respectively; (3) the degree of distortion of the BH shadow is different for various quantum correction parameters $m$; with the increase of the values of $m$, the shadow will become more and more obvious; (4) the results of different BH type parameter $n$ differ greatly. Since the value of $m$ in actual physics should be very small, the current observations of EHT cannot distinguish quantum effect from BH shadow, and can only constrain the upper limit of $m$. In future BH shadow measurements, it will be possible to distinguish quantum deformed black holes, which will help to better understand the quantum effects of BHs.}


\begin{document}
\maketitle
\flushbottom

\section{Introduction}
\label{intro}

Black hole (BH) physics is a hot field in physics and astronomy at present. Its main problem is how to understand quantum effects near the event horizon of BH. These problems involve quantization of gravity and are therefore very complex \cite{1998bhp..book.....F}. With the emergence of observational evidence for BH, these problems have become more prominent in recent years \cite{2016PhRvL.116f1102A,2019ApJ...875L...1E}. Of course, scientists are very interested in how to test the related theoretical models from practical perspective. So far, the most direct way to detect BH is to make measurements with the BH Event Horizon Telescope (EHT)\cite{2019ApJ...875L...1E}. EHT is based on very long baseline interferometry (VLBI) to measure BH shadow. It focused on the supermassive BH at the center of the Milky Way and the M87 supermassive BH. On April 10, 2019, it released observations of the M87 supermassive BH, which was found to have a mass of $6.5\times 10^{9}M_{\odot}$. Since this event, a great deal of research has been done on general relativity using the M87 data \cite{2019ApJ...875L...1E,2019ApJ...875L...2E,2019ApJ...875L...3E,2019ApJ...875L...4E,2019ApJ...875L...5E,2019ApJ...875L...6E,2021ApJ...910L..12E,2021ApJ...910L..13E}. 

The basic picture of the BH shadow is that when light passes near a BH and then reaches earth, the strong gravitational field of the BH means that the background photons only reach earth outside of a certain area. The BH shadow is then observed, and its shape and size depend on the parameters of the central BH; So scientists can infer the presence of a BH by measuring its shadow, and use that as a basis for studying the strong gravitational field \cite{2019LRR....22....4C,2019PhRvD.100d4057B,2019PhRvD.100b4018G,2020PhRvL.125n1104P,2020JCAP...09..026K,2019JCAP...08..030W,2019Univ....5..220C,2019GReGr..51..137P,2019arXiv191212629L}. 

The shadow of Schwarzschild BH was first calculated by Synge \cite{1966MNRAS.131..463S}. Since then, using analytic method to calculate the BH shadow has become a hot topic \cite{1966MNRAS.131..463S,1979A&A....75..228L,2005PASJ...57..273T,2018PhRvD..97f4021T,2013Ap&SS.344..429A,2012PhRvD..85f4019A,2013PhRvD..87d4057A,2013PhRvD..88f4004A,2016PhRvD..94h4025Y,2017PhLB..768..373C,2016arXiv161009477D,2016PhRvD..93j4004A,2018PhRvD..97j4062P,2017arXiv171209793K,2014PhRvD..90b4073P,2018EPJC...78..399A,2015PhRvL.115u1102C,2018PhRvD..97h4024G,2012PhRvD..86j3001Y,2018PhRvD..98d4053C,2018GReGr..50...42C}. The accretion process of BH can affect the image of BH, so it is necessary to study this process by analytical method and numerical simulation \cite{2000ApJ...528L..13F,2007CQGra..24S.259N,2010ApJ...717.1092D,2014A&A...570A...7M,2015ApJ...799....1C,2016ApJ...820..137B,2017ApJ...837..180G}. Xu et al., 2018, calculated the effect of the distribution of the dark matter (DM) halo at the center of the Milky Way on the BH shadow \cite{2018JCAP...10..046X,2018JCAP...07..015H,2018JCAP...12..040H}. They found that if the resolution of EHT is further improved, it is possible to detect the DM effect near the Sgr A* BH and distinguish the state properties of ambient matter near the BH.  

Now let's discuss the interesting question of whether it is possible to use future EHT measurements to study the quantum effect near a BH. We discuss this problem in the classical case by looking for modified forms of space-time of the geometric BH under quantum effect, and then calculating the possible observational effects of these modified BH metrics. Kazakov et al., began this research in 1994 by studying the quantum modification of the Schwarzschild BH \cite{1994NuPhB.429..153K}. Based on this approach, scientists then calculated the potential function satisfied by the quantum field in the space-time background of the Schwarzschild BH \cite{2016NuPhB.909..173A}. In the reference \cite{2017EPJC...77..243C}, Calmet et al., calculated the quantum modification of the Schwarzschild BH in general relativity based on the effective field theory method. Good et al., proposed a Schwarzschild BH solution for the BH evaporation model, in which the BH information is not lost \cite{2020arXiv200301333G}. Recently, Berry et al., obtained the universal form of the quantum deformed BH metric based on previous work \cite{2021arXiv210202471B}. Xu et al., extended the BH metric to the case of the rotating BH \cite{2021arXiv210913813}. Based on these results, we will calculate the shadow of the rotating quantum-deformed BH, thus proposing the possibility of using EHT to test the quantum deformed BH.    

The basic structure of this article is as follows. In section 2, we introduce the spherically symmetric and axisymmetric quantum deformed BHs. In section 3, the geodesic equations of light rays are derived analytically and the general analytical solutions are obtained. In section 4, we calculate the change of the quantum deformation parameters to the shape of the BH shadow. In section 5, we calculate the change of the quantum deformation parameters to the energy emission rate. Finally, summarize the whole article.

\section{Quantum deformed BH}
\label{2}

\subsection{Spherically symmetric quantum deformed BH}

The quantum deformed BH introduced here is mainly related to the theory of the two-dimensional dilaton gravity \cite{1994NuPhB.429..153K,1992NuPhB.382..259R,2021arXiv210202471B}. In this theory of gravity, the action $S$ is

\begin{equation}
S=-\dfrac{1}{8}\int d^{2}z\sqrt{-g}\left[r^{2}R^{(2)}-2(\nabla r)^{2}+\dfrac{2}{\kappa}U(r)\right],
\label{2.1}
\end{equation}

where $g$ is metric determinant, $R^{(2)}$ is two-dimensional Ricci scalar, $\kappa$ is dimensionless constant, and $U(r)$ is dilaton potential energy. By variational calculation of the action (\ref{2.1}), the corresponding field equation can be obtained. And then, by solving the field equation, we can get the space-time metric  

\begin{equation}
ds^{2}=-f(r)dt^{2}+\dfrac{1}{f(r)}dr^{2}+r^{2}\left(d\theta^{2}+\sin^{2}\theta d\phi^{2}\right),
\label{2.2}
\end{equation}

where, the mathematical relation between the metric coefficient function $f(r)$ and the dilaton potential energy is

\begin{equation}
f(r)=\dfrac{1}{r}\int^{r}U(\rho)d\rho-\dfrac{2M}{r},
\label{2.3}
\end{equation}

By selecting the form of the potential function $U(r)$, the modified form of the spherically symmetric BH in two-dimensional dilaton gravity theory can be obtained. For example, a special quantum modified BH, Kazakov and Solodukhin chose the potential function $U(r)$ as the following form \cite{1994NuPhB.429..153K}

\begin{equation}
U(r)=\dfrac{r}{\sqrt{r^{2}-m^{2}}},
\label{2.4}
\end{equation}

In equations (\ref{2.2}) and (\ref{2.4}), $M$ is the mass of the spherically symmetric BH, and its physical meaning is the total mass of the system when the radial distance is infinite; $m$ is quantum deformation parameter, which represents the modification of the spherically symmetric metric by quantum effect after considering the two-dimensional dilaton gravity theory.

According to the two-dimensional dilaton theory of gravity, as long as the potential function $U(r)$ satisfying the modified gravitational field equation can be used as a suitable choice. On this basis, Berry, Simpson and Visser generalized the space-time metric to the following form \cite{2021arXiv210202471B}

\begin{equation}
f(r)=f_{n}(r)=\left(1-\dfrac{m^{2}}{r^{2}}\right)^{\frac{n}{2}}-\dfrac{2M}{r}.
\label{2.5}
\end{equation}

where, the quantum deformation parameter $m\geq 0$, $n=0, 1, 3, 5...$; It is worth noting that when $n=0$ or $m=0$, equation (\ref{2.5}) reduces to the case of the Schwarzschild BH. In addition, for the general parameter selection, the space-time metric is the modified BH space-time. For example, when $n=1$, the space-time metric degrades to the case of Kazakov-solodukhin \cite{1994NuPhB.429..153K}, the quantum modified BH exhibits a series of special properties that can be analytically calculated. When $n=3$ or $5$, the space-time metric will change into Christoffel-symbol-regular and Curvature-regular space-time, and this space-time metric provides a basis for us to analyze the influences of quantum effect on Schwarzschild BH.

\subsection{Axisymmetric quantum deformed BH}

On the basis of the spherically symmetric quantum deformed BH (equations (\ref{2.2}) and (\ref{2.5})), Xu and Tang generalized it to the case of the axisymmetric BH using the Newman-Janis method based on complex transformation \cite{2021arXiv210913813}. If the spherically symmetric metric satisfies $f(r)=g(r)$ (where $g(r)$ is $g_{rr}^{-1}$), regardless of the form of $h(r)$, Einstein gravitational field equations can be solved by solving partial differential equations. Through more complicated derivation, they found that the metric of the quantum deformed BH under rotational symmetry is  

\begin{equation}
ds^{2}=-\left[1-\dfrac{r^{2}+2Mr-r^{2}\left(1-\dfrac{m^{2}}{r^{2}}\right)^{\frac{n}{2}}}{{\Sigma}^{2}}\right]dt^{2}+\dfrac{\Sigma^{2}}{\Delta}dr^{2}$$$$
-\dfrac{2a\sin^{2}\theta d\phi dt}{\Sigma^{2}}\left[r^{2}+2Mr-r^{2}\left(1-\dfrac{m^{2}}{r^{2}}\right)^{\frac{n}{2}}\right]$$$$
+\Sigma^{2}d\theta^{2}+\dfrac{\sin^{2}\theta}{\Sigma^{2}}\left[\left(r^{2}+a^{2}\right)^{2}-a^{2}\Delta\sin^{2}\theta \right]d\phi^{2} ,
\label{2.6}
\end{equation}

where, the unknown functions in the metric are expressed as follows:

\begin{equation}
\begin{aligned}
& \Sigma^{2}=r^{2}+a^{2}\cos^{2}\theta,    \\
& \Delta=r^{2}\left(1-\dfrac{m^{2}}{r^{2}}\right)^{\frac{n}{2}}-2Mr+a^{2}.
\end{aligned}
\label{2.7}
\end{equation}

As mentioned earlier, when the quantum deformation parameter $m=0$ (or $n=0$), the axisymmetric quantum deformed BH (equation (\ref{2.6})) degenerates into Kerr BH; When $n=1$, equation (\ref{2.6}) degrades into a rotational promotion of the case of Kazakov-Solodukhin, and the BH spin can be increased to extreme situation. For other values of $n$ and $m$, corresponding to various types of axisymmetric quantum deformed BH \cite{2021arXiv210913813}.  
  
From this axisymmetric quantum deformed BH, it is possible to calculate the influences of the quantum effect on various BH related observations.

\section{Geodesic equations under axisymmetric BH}
\label{3}

Based on the metric equation (\ref{2.6}) of the axisymmetric BH, we calculate the geodesic motion of a test particle in space-time. Here, we will deal with it according to Hamilton-Jacobi equation and Carter variable separation method \cite{1968PhRv..174.1559C}. In the general space-time background, the form of the Hamilton-Jacobi equation is as follows:

\begin{equation}
\dfrac{\partial S}{\partial \eta}=-\dfrac{1}{2}g^{\mu\nu}\dfrac{\partial S}{\partial x^{\mu}}\dfrac{\partial S}{\partial x^{\nu}},
\label{3.1}
\end{equation}

Where $\eta$ is affine parameter on geodesic, $S$ is Jacobi action. In Carter variable separation method, Jacobi action $S$ can be written as  

\begin{equation}
S=\dfrac{1}{2}\bar{m}^{2}\eta-Et+L\phi+S_{r}(r)+S_{\theta}(\theta),
\label{3.2}
\end{equation}

Where $\bar{m}$ is the mass of the test particle, and $E$ and $L$ are the energy and angular momentum of the particle respectively. $S_{r}(r)$ is a radial unknown function and $S_{\theta}(\theta)$ is an angular unknown function. Substituting the Jacobi action equation (\ref{3.2}) into the Hamilton-Jacobi equation (\ref{3.1}), by simple derivation, we can obtain motion equations, namely the geodesic equations, satisfied by the test particle in the space-time of the quantum deformed BH. In BH physics, geodesic equations can be expressed as \cite{Chandrasekhar:1985kt}: 

\begin{equation}
\Sigma^{2}\dfrac{dt}{d\eta}=\dfrac{r^{2}+a^{2}}{\Delta}\left[E \left(r^{2}+a^{2} \right)-aL \right]-a \left(aE\sin^{2}\theta-L \right),
\label{3.3}
\end{equation}

\begin{equation}
\Sigma^{2}\dfrac{dr}{d\eta}=\sqrt{R},
\label{3.4}
\end{equation}

\begin{equation}
\Sigma^{2}\dfrac{d\theta}{d\eta}=\sqrt{\circled{H}},
\label{3.5}
\end{equation}

\begin{equation}
\Sigma^{2}\dfrac{d\phi}{d\eta}=\dfrac{a}{\Delta}\left[E\left(r^{2}+a^{2}\right)-aL \right]-\left(aE-\dfrac{L}{\sin^{2}\theta}\right),
\label{3.6}
\end{equation}

In the geodesic equations (\ref{3.3}) $\sim$ (\ref{3.6}), the basic expressions of the newly introduced functions $R$ and $\circled{H}$ are 

\begin{equation}
R=\left[E\left(r^{2}+a^{2}\right)-aL\right]^{2}-\Delta \left[m^{2}r^{2}+(aE-L)^{2}+K \right],
\label{3.7}
\end{equation}

\begin{equation}
\circled{H}=K-\left(\dfrac{L^{2}}{\sin^{2}\theta}-a^{2}E^{2}\right)\cos^{2}\theta.
\label{3.8}
\end{equation}

In the above two equations, $K$ is Carter constant. Equations (\ref{3.3}) $\sim$ (\ref{3.8}) fully describe the motion of the test particle. Given an angle $\theta$, the geodesic equations for the plane $\theta$ can be solved. In this work, we are interested in the geodesic motion of photons near the event horizon of BH, and the test particle has a rest mass of $0$. Near BH, the boundary of the photon sphere and its shape are determined by the unstable orbit of the particle. Therefore, the following conditions must be satisfied:  

\begin{equation}
R=0,
\label{3.9}
\end{equation}

\begin{equation}
\dfrac{\partial R}{\partial r}=0,
\label{3.10}
\end{equation}

For an observer at infinity on the equatorial plane of the BH space-time, the photon geodesic satisfying the equations (\ref{3.9}) and (\ref{3.10}) can be expressed by two parameters, which are defined as:

\begin{equation}
\xi=\dfrac{L}{E},
\label{3.11}
\end{equation}

\begin{equation}
\bar{\eta}=\dfrac{K}{E^{2}},
\label{3.12}
\end{equation}

Combining equations (\ref{3.9}) $\sim$ (\ref{3.12}) and substituting the expression of $R$, the differential equations satisfied by parameters $\xi$ and $\bar{\eta}$ can be obtained  

\begin{equation}
\left[\bar{\eta}+(\xi-a)^{2}\right]\left(r^{2}f(r)+a^{2}\right)-\left(r^{2}+a^{2}-a\xi\right)^{2}=0,
\label{3.13}
\end{equation}

\begin{equation}
\left[\bar{\eta}+(\xi-a)^{2}\right]\left(2rf(r)+r^{2}f^{'}(r)\right)-4r\left(r^{2}+a^{2}-a\xi\right)=0,
\label{3.14}
\end{equation}

The solution of the above equations is

\begin{equation}
\xi=\dfrac{\left(r^{2}+a^{2}\right)\left(rf^{'}(r)+2f(r)\right)-4\left(r^{2}f(r)+a^{2}\right)}{a \left(rf^{'}(r)+2f(r)\right)},
\label{3.15}
\end{equation}

\begin{equation}
\bar{\eta}=\dfrac{r^{2}\left[8a^{2}f^{'}(r)-r\left(2f(r)-rf^{'}(r)\right)^{2}\right]}{a^{2}\left(rf^{'}(r)+2f(r)\right)^{2}}.
\label{3.16}
\end{equation}

According to equation (\ref{2.5}), the expression of $f^{'}(r)=f_{n}^{'}(r)$ is

\begin{equation}
f^{'}(r)=f_{n}^{'}(r)=\dfrac{2M}{r^{2}}+\dfrac{nm^{2}}{r^{3}}\left(1-\dfrac{m^{2}}{r^{2}} \right)^{\frac{n-2}{2}}.
\label{3.17}
\end{equation}

\section{The shadow of quantum deformed BH}
\label{4}

\subsection{The shape of shadow}

In the previous section, we calculated the geodesic equations of the photon and the shape of the photon sphere, but these quantities are not the physical quantities measured by an observer on Earth. Compared to BH, an observer on Earth can be approximated as an observer at infinity; Therefore, for observers on earth, the celestial coordinate system can be introduced, whose components $\alpha$ and $\beta$ are defined as 

\begin{equation}
\alpha=\lim\limits_{r_{0}\to\infty}\left[-r_{0}^{2}\sin\theta_{0}\dfrac{d\phi}{dr} \right],
\label{4.1}
\end{equation}

\begin{equation}
\beta=\lim\limits_{r_{0}\to\infty}\left[r_{0}^{2}\dfrac{d\theta}{dr} \right],
\label{4.2}
\end{equation}

Where, $r_{0}$ is the distance between the observer and BH, that is, the distance between the earth and BH; $\theta_{0}$ is the angle between the observer's line of sight and the spin axis of BH; $\alpha$ and $\beta$ are the projections of the BH shadow on the celestial coordinate system. Substituting the photon geodesic equations (\ref{3.3}) $\sim$ (\ref{3.6}) into equations (\ref{4.1}) and (\ref{4.2}), we get the following relations

\begin{equation}
\alpha=-\dfrac{\xi}{\sin\theta}\mid_{\theta=\frac{\pi}{2}}=-\xi,
\label{4.3}
\end{equation}

\begin{equation}
\beta=\pm\sqrt{\bar{\eta}+a^{2}\cos^{2}\theta-\xi^{2}\cot^{2}\theta}\mid_{\theta=\frac{\pi}{2}}=\pm\sqrt{\bar{\eta}}.
\label{4.4}
\end{equation}

\begin{figure}[htbp]
  \centering
   \includegraphics[scale=0.29]{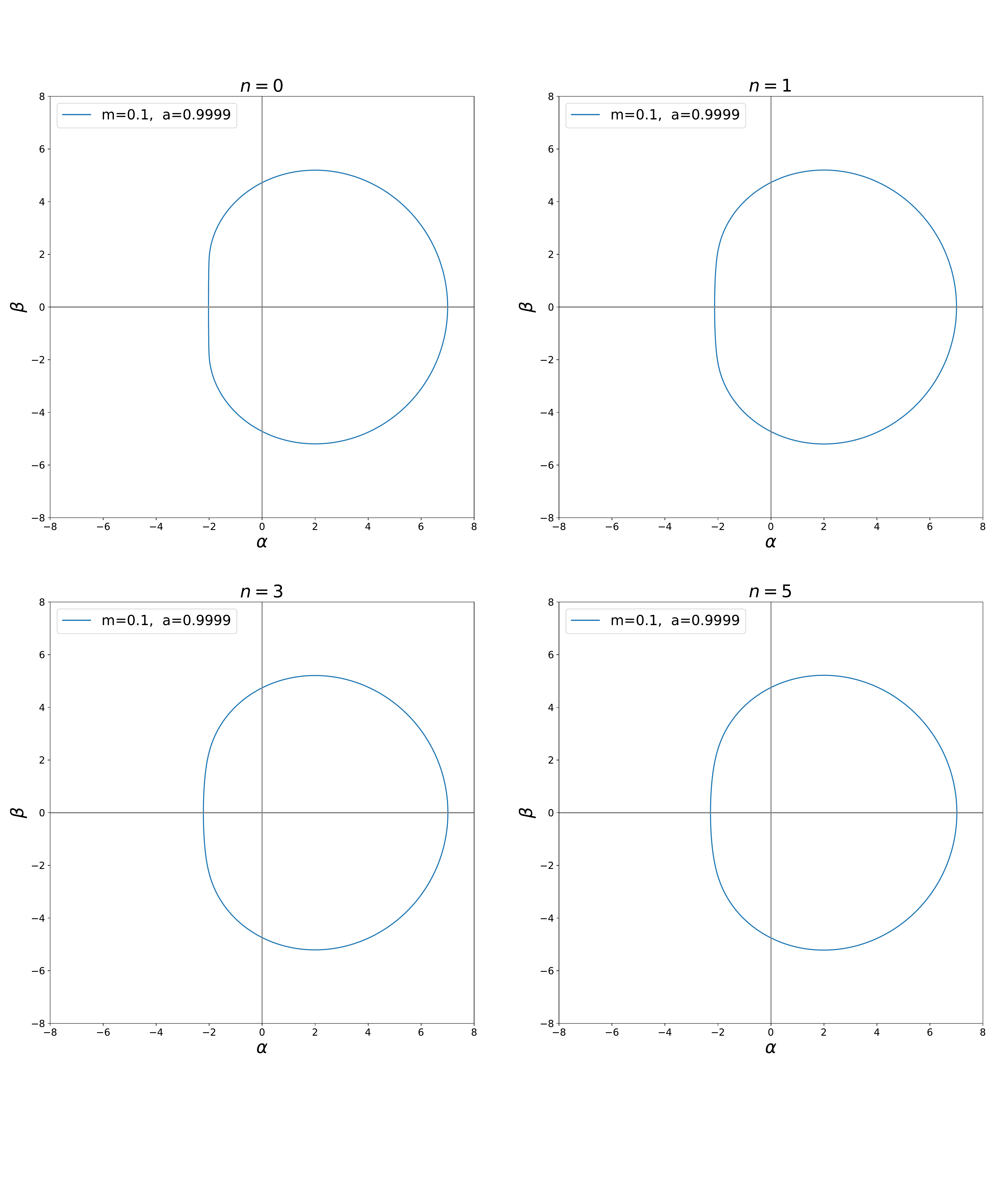}
   \caption{Shape of BH shadow under various types of quantum deformed BHs ($n=0, 1, 3, 5$). Here, the quantum deformation parameter $m=0.1$ and the BH spin $a=0.9999$.}
  \label{shadow_n_1}
\end{figure}

\begin{figure}[htbp]
  \centering
   \includegraphics[scale=0.29]{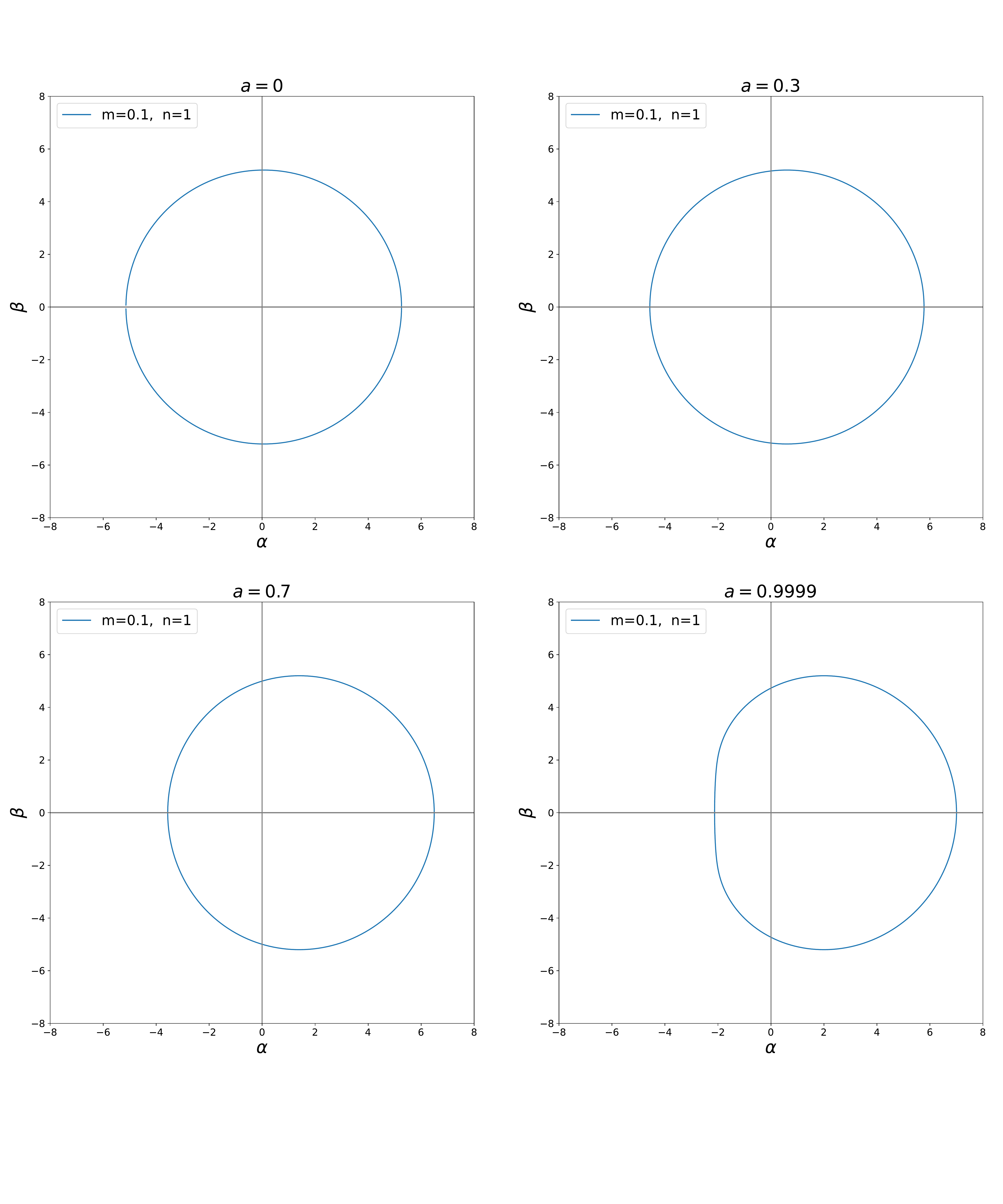}
   \caption{Shape of BH shadow changed by different BH spin ($a=0, 0.3, 0.7, 0.9999$). Here, the quantum deformation parameter $m=0.1$ and the BH model parameter $n=1$. }
  \label{shadow_a_1}
\end{figure}

\begin{figure}[htbp]
  \centering
   \includegraphics[scale=0.29]{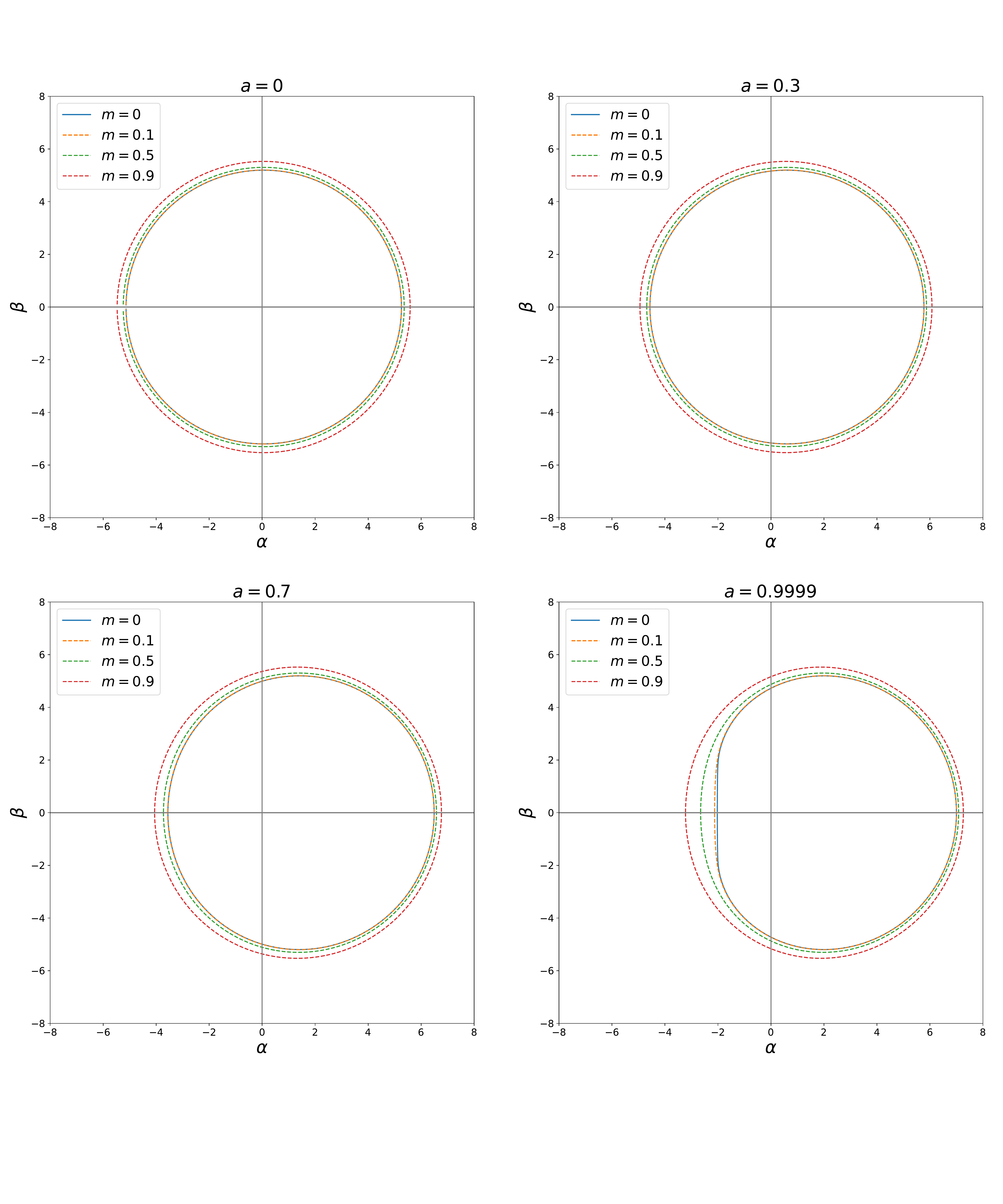}
   \caption{The shape of the Kerr-BH ($n=1$) shadow changes with different spin ($a=0, 0.3, 0.7, 0.9999$) and quantum deformation parameters ($m=0, 0.1, 0.5, 0.9$).}
  \label{shadow_ma_1}
\end{figure}

Using equations (\ref{3.15}), (\ref{3.16}), (\ref{4.3}) and (\ref{4.4}), we draw the shape of the BH shadow, as shown in Figures \ref{shadow_n_1}, \ref{shadow_a_1} and \ref{shadow_ma_1}, and the main results are as follows:    

(1) According to the results of numerical calculation (Figure \ref{shadow_n_1}), the BH shadows of various types of quantum deformed BH are very different. When $n=0$, the BH metric equation (\ref{2.6}) degenerates to the case of the Kerr BH, whose BH shadow is exactly the same as that of the Kerr BH. When $n=1$, $3$, or $5$, the scale of the shadow of the quantum deformed BH increases with the increase of $n$, which indicates that the Kerr BH ($n=0$) has the smallest shadow and the curvature BH ($n=5$) has the largest shadow.

(2) When the spin of BH is $0$, the BH shadow of the quantum deformed BH is a standard circle. As spin $a$ grows, the shape of the BH shadow becomes more and more distorted; When $a\rightarrow 1$, the distortion of the BH shadow shape reaches the maximum. These results are consistent with the case of the Kerr BH, except that the modification of the BH shadow by quantum effect is considered here.

(3) The numerical results of the influence of the different quantum deformation parameters $m$ on the shape of shadow are shown in Figure \ref{shadow_ma_1}, indicating that when the quantum deformation parameter is not considered ($m=0$), the results are completely consistent with that of the Kerr BH; With the increasing of the quantum deformation parameter $m$, the scale of shadow will also increase. From the quantum deformed BH, we can know that the value of $m$ will be very small, so our discussion here is only a theoretical model, and has no actual observational effect.

For the shadow shape of the quantum deformed BH, the modification parameters $m$ and $n$ of quantum effect have obvious changes to it. Using these results, it will be possible to examine whether the shape changes of the BH shadows are caused by quantum effect and to study the details of quantum effect of BH through future observations in the EHT (further improve resolution) project.

\subsection{The scale of shadow}

In the previous part, we only studied the BH shadow from the shape, although also quantitative, but the description of the BH shadow is not detailed enough. Here, we will use two observational parameters to describe the shape of the BH shadow, which are BH shadow radius $R_{s}$ and distortion parameter $\delta_{s}$ respectively \cite{2009PhRvD..80b4042H,2018JCAP...07..015H,2018JCAP...10..046X}. The shadow radius $R_{s}$ can be defined as the radius of the circle determined by $A(\alpha_{r}, 0)$, $B(\alpha_{t}, \beta_{t})$ and $D(\alpha_{b}, \beta_{b})$; The distortion parameter $\delta_{s}$ can be defined as the distortion degree of the BH shadow relative to the standard circle. $R_{s}$ and $\delta_{s}$ can basically describe the details of the scale and shape of the BH shadow. According to the definition, $R_{s}$ and $\delta_{s}$ can be expressed as: 

\begin{equation}
R_{s}=\dfrac{\beta_{t}^{2}+(\alpha_{t}-\alpha_{r})^{2}}{2|\alpha_{r}-\alpha_{t}|}
\label{4.5},
\end{equation}

\begin{equation}
\delta_{s}=\dfrac{d_{s}}{R_{s}}=\dfrac{|\alpha_{p}-\tilde{\alpha}_{p}|}{R_{s}}.
\label{4.6}
\end{equation}

where $d_{s}$ is the distance between two points $C(\alpha_{p}, 0)$ and $F(\bar{\alpha}_{p}, 0)$, $C(\alpha_{p}, 0)$ is the left intersection of the BH shadow and the $\alpha$ axis, and $F(\bar{\alpha}_{p}, 0)$ is the left intersection of the reference circle and the $\alpha$ axis. Interestingly, the shape of the shadow of a BH with zero spin is a standard circle, without any distortion in shape.

\begin{figure}[htbp]
  \centering
   \includegraphics[scale=0.39]{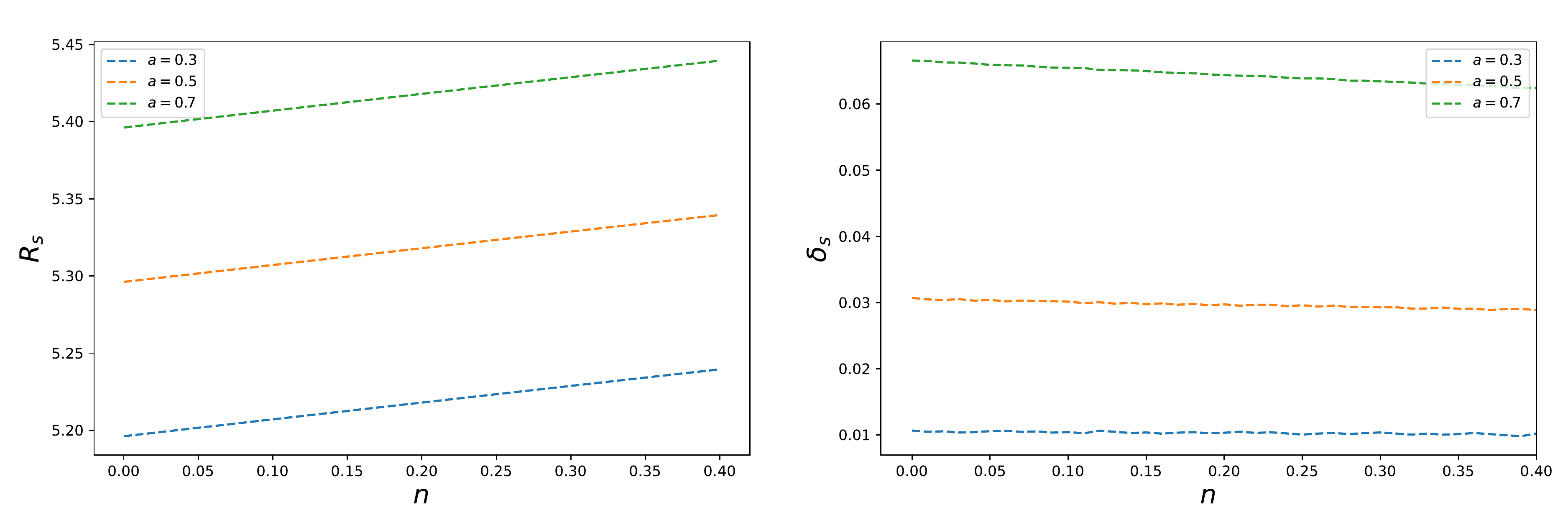}
   \caption{The radius $R_{s}$ (left) and the distortion parameter $\delta_{s}$ (right) of the BH shadow vary with the model parameter $n$ under different BH spin ($a=0.3, 0.5, 0.7$). Where the quantum deformation parameter $m=0.5$.}
  \label{Rs_deltaS_n_1}
\end{figure}

\begin{figure}[htbp]
  \centering
   \includegraphics[scale=0.39]{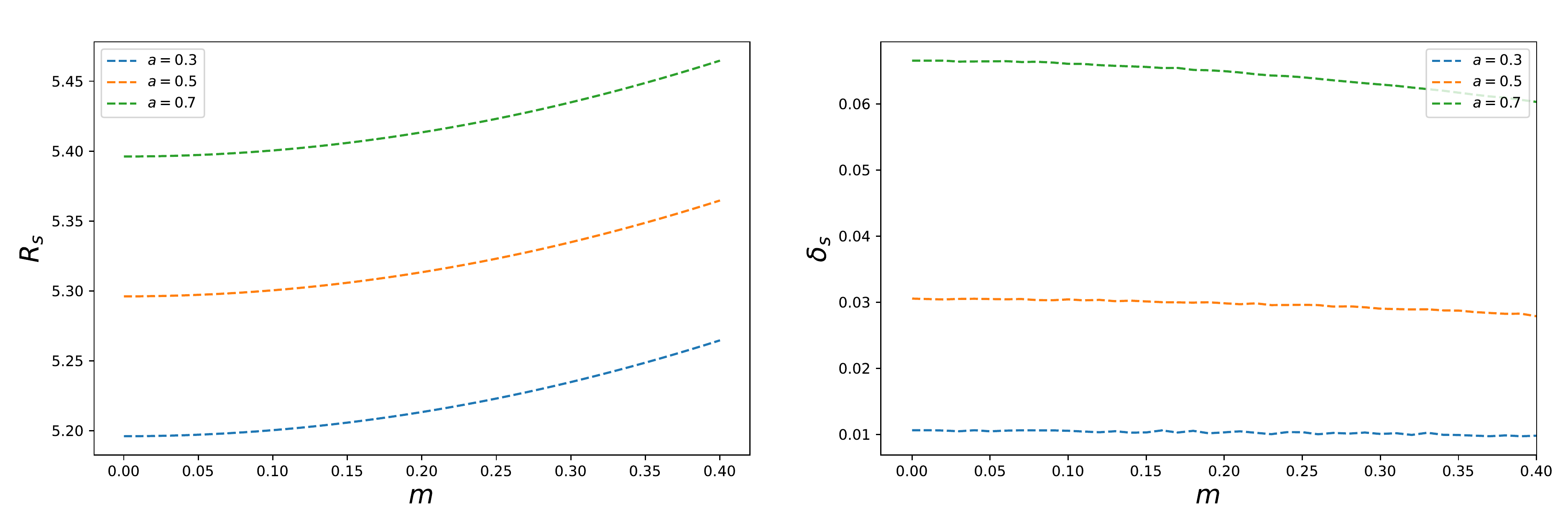}
   \caption{The radius $R_{s}$ (left) and the distortion parameter $\delta_{s}$ (right) of the BH shadow vary with the quantum deformation parameter $m$ under different BH spin ($a=0.3, 0.5, 0.7$). Where the model parameter $n=1$.}
  \label{Rs_deltaS_m_1}
\end{figure}

\begin{figure}[htbp]
  \centering
   \includegraphics[scale=0.39]{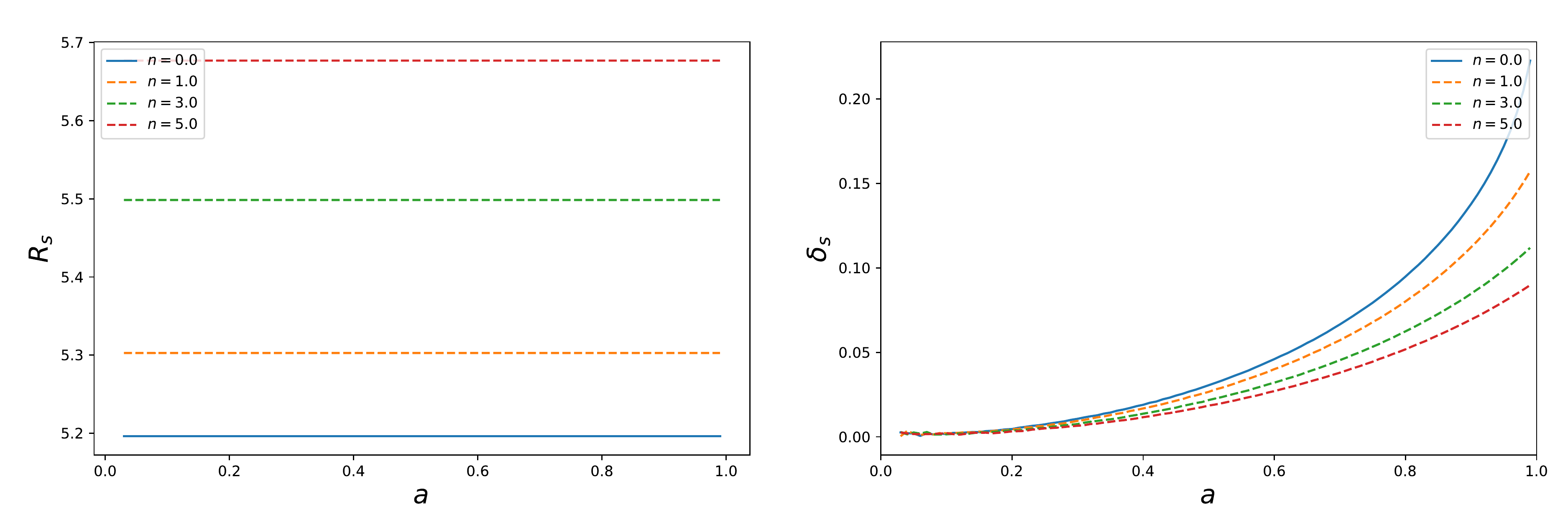}
   \caption{The radius $R_{s}$ (left) and the distortion parameter $\delta_{s}$ (right) of the BH shadow vary with BH spin $a$ under different model parameters ($n=0, 1, 3, 5$). Where the quantum deformation parameter $m=0.5$.}
  \label{Rs_deltaS_a_1}
\end{figure}

Figures \ref{Rs_deltaS_n_1}-\ref{Rs_deltaS_a_1} show the changes of the BH shadow radius $R_{s}$ and the distortion parameter $\delta_{s}$ with the model parameters. By analyzing the numerical results, we find that: (1) With the increase of the BH type parameter $n$, that is, from the Schwarzchild BH case, to the metric-regular case, to the Christoffel-symbol-regular case, and then to the Curvature-regular case, $R_{s}$ and $\delta_{s}$ show monotone increasing and decreasing functions, respectively; (2) When $n=1$, that is, considering the metric-regular case, $R_{s}$ is the increasing function of the quantum deformation parameter $m$, and $\delta_{s}$ is the decreasing function of $m$; (3) For an ordinary quantum deformed BH, i.e. $m\neq 0$, we find that $R_{s}$ and $\delta_{s}$ both increase with the spin of BH.

\section{Energy emission rate}
\label{5}

When the observer is at infinity, the size of the BH shadow is proportional to the high energy absorption cross section of the particle, which is similar to taking the BH shadow as a black body. Scientists found that the high energy absorption cross section ($\sigma_{eim}$) oscillates around a constant. For a spherically symmetric quantum deformed BH, the geometric cross sections of $\sigma_{eim}$ and the photon sphere are approximately equal, i.e., $\sigma_{eim}\approx \pi R_{s}^{2}$ \cite{1973PhRvD...7.2807M,2013JCAP...11..063W}, Where $R_{s}$ is the radius of the reference circle defined in the previous section. In our work, as shown in Figures \ref{shadow_n_1}-\ref{shadow_ma_1}, except for extreme and near-extreme Kerr BH, the shadow shapes of other BH are close to the circle, so it is approximately reasonable to apply the results of spherically symmetric BH to axisymmetric BH. Therefore, for a general axisymmetric quantum deformed BH, its energy emission rate is  

\begin{equation}
\dfrac{d^{2}E(\omega)}{d\omega dt}=\dfrac{2\pi^{2}\omega^{3}\sigma_{eim}}{e^{\frac{\omega}{T}}-1},
\label{5.1}
\end{equation}

Here, $T$ is the Hawking temperature corresponding to the event horizon of the quantum deformed BH, and $\omega$ is the frequency of photon. According to the definition of BH temperature, its expression is as follows 

\begin{equation}
T=\lim\limits_{\theta \to 0, r \to r_{+}}\dfrac{1}{2\pi\sqrt{g_{rr}}}\dfrac{\partial \sqrt{g_{tt}}}{\partial r},
\label{5.2}
\end{equation}

The specific expressions of $g_{tt}$ and $g_{rr}$ are shown in equations (\ref{2.6}) and (\ref{2.7}). By substituting the metric coefficients into equation (\ref{5.2}), we can obtain the general form of Hawking temperature  

\begin{equation}
T=\dfrac{2a^{2}r_{+}[f(r_{+})-1]+r_{+}^{2}f^{'}(r_{+})\left(r_{+}^{2}+a^{2}\right)}{4\pi^{2}\left(r_{+}^{2}+a^{2}\right)^{2}}.
\label{5.3}
\end{equation}

\begin{figure}[htbp]
  \centering
   \includegraphics[scale=0.39]{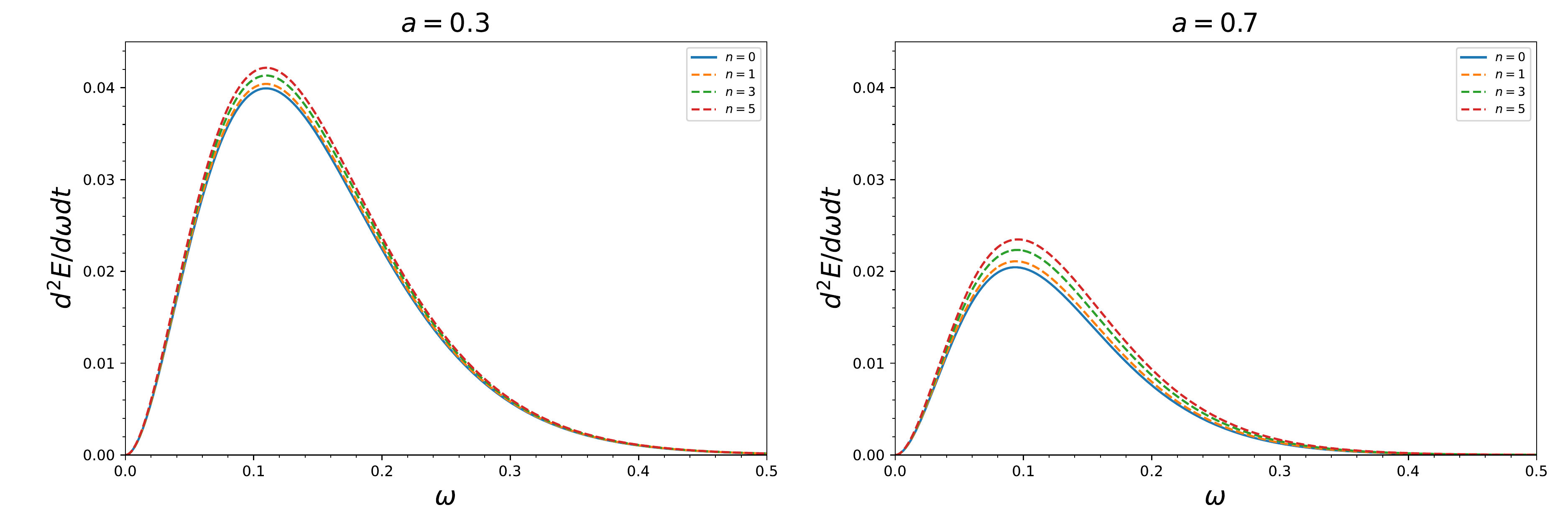}
   \caption{The energy emission rate of the BH evolves with the frequency $\omega$ under different values of the BH spin ($a=0.3$ (left) and $a=0.7$ (right)) and the model parameter ($n=0, 1, 3, 5$). Where the quantum deformation parameter $m=0.2$.}
  \label{emiss_rate_n_1_1}
\end{figure}

\begin{figure}[htbp]
  \centering
   \includegraphics[scale=0.39]{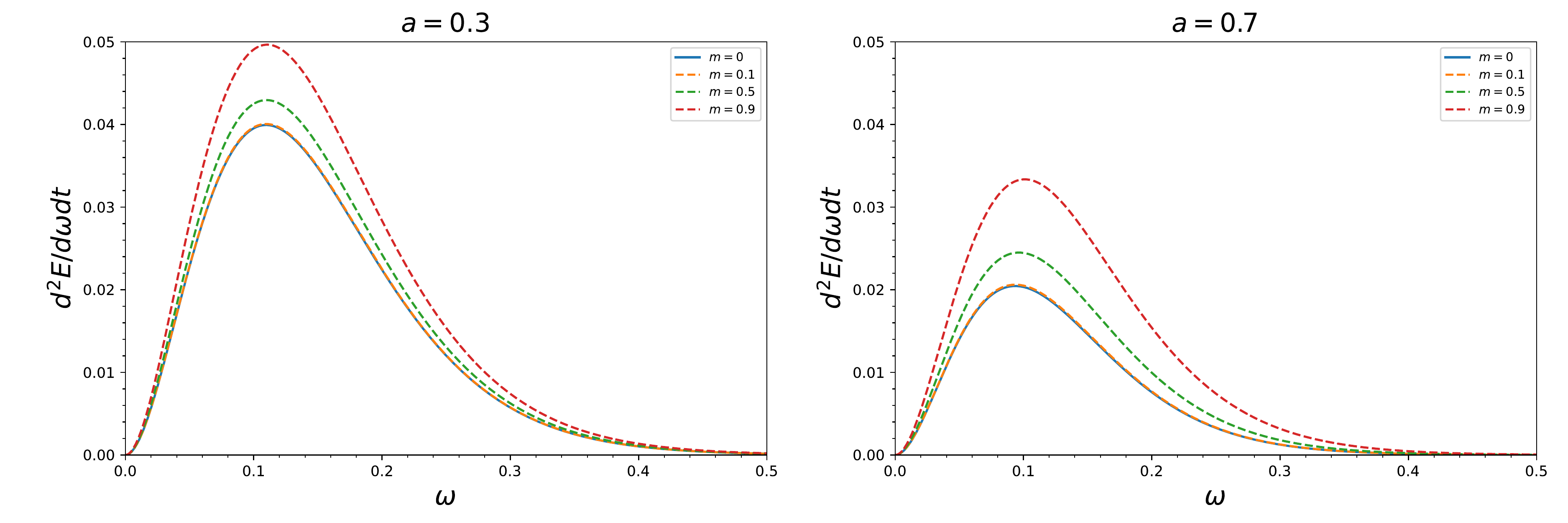}
   \caption{The energy emission rate of the BH evolves with the frequency $\omega$ under different values of the BH spin ($a=0.3$ (left) and $a=0.7$ (right)) and the quantum deformation parameter ($m=0, 0.1, 0.5, 0.9$). Where the model parameter $n=1$.}
  \label{emiss_rate_m_1_1}
\end{figure}

Numerical calculation is performed for equation (\ref{5.1}), as shown in Figures \ref{emiss_rate_n_1_1} and \ref{emiss_rate_m_1_1}. The main results are as follows: (1) the energy emission rate of various types of quantum deformed BH ($n=0, 1, 3, 5$) decreases with the increase of $n$ value; (2) the energy emission rate of the quantum deformed BH decreases with the increase of the quantum deformation parameter $m$, that is to say, the more significant the quantum effect is, the lower the energy emission rate is; (3) for a particular quantum deformed BH, when the spin is $0$, the energy emission rate is the largest, and the energy emission rate decreases with the increase of the spin.

\section{Summary}
\label{6}

In this work, we study the properties of the shadow of quantum deformed BH and discuss the possibility of using EHT to test the quantum properties of BH. An observer at infinity (an observer on Earth can be approximated as an observer at infinity) would see that the shadow of a spherically symmetric BH is a circle whose size is closely related to the parameters of the quantum deformed BH; If the BH is rotating, then the shape of the shadow will depend on the metric structure of the quantum deformed BH, which is the shape of the circle after being distorted. We find that the structure of the shadow depends entirely on the parameters of the quantum deformed BH, namely, the BH type parameter $n$, the quantum deformed parameter $m$ and the BH spin $a$. 

Based on these results, we believe that if the resolution of EHT is further improved, it will be possible to distinguish the quantum effect of BH in EHT measurements, thus providing a new way to test the quantum effect of BH by observation. This paper is the theoretical calculation of this idea.

\acknowledgments
We acknowledge the anonymous referee for a constructive report that has significantly improved this paper. We acknowledge the  Special Natural Science Fund of Guizhou University (grant
No. X2020068) and the financial support from the China Postdoctoral Science Foundation funded project under grants No. 2019M650846.

\end{document}